\documentclass[conference]{IEEEtran}
\IEEEoverridecommandlockouts
\usepackage{cite}
\usepackage{amsmath,amssymb,amsfonts}
\usepackage{algorithmic}
\usepackage{graphicx}
\usepackage{textcomp}
\usepackage{xcolor}
\usepackage{tabularx}
\usepackage{caption}
\usepackage{url}
\usepackage{booktabs}
\def\BibTeX{{\rm B\kern-.05em{\sc i\kern-.025em b}\kern-.08em
T\kern-.1667em\lower.7ex\hbox{E}\kern-.125emX}}

\makeatletter
\newcommand{\linebreakand}{%
 \end{@IEEEauthorhalign}
 \hfill\mbox{}\par
 \mbox{}\hfill\begin{@IEEEauthorhalign}
\newcommand{\customIEEEauthorrefmark}[1]{\textsuperscript{a}}

\newcommand{\customIEEEauthorrefmark}[2]{\textsuperscript{b}}

\newcommand{\customIEEEauthorrefmark}[3]{\textsuperscript{c}}

\newcommand{\customIEEEauthorrefmark}[4]{\textsuperscript{d}}
\newcommand{\customIEEEauthorrefmark}[5]{\textsuperscript{*}}
}
\makeatother

\begin{document}

\title{A Multi-Resolution Mutual Learning Network for Multi-Label ECG Classification\\
}
\author{
    \IEEEauthorblockN{Wei Huang\IEEEauthorrefmark{2,3}, Ning Wang\IEEEauthorrefmark{2,3,}\IEEEauthorrefmark{1}, Panpan Feng\IEEEauthorrefmark{4}, Haiyan Wang\IEEEauthorrefmark{5},  Zongmin Wang\IEEEauthorrefmark{2,3,}\IEEEauthorrefmark{1}
   and Bing Zhou\IEEEauthorrefmark{2,3}
    }
    \IEEEauthorblockA{\IEEEauthorrefmark{2}\textit{School of Computer and Artificial Intelligence, Zhengzhou University, Zhengzhou, China}}
    \IEEEauthorblockA{\IEEEauthorrefmark{3}\textit{Cooperative Innovation Center of Internet Healthcare, Zhengzhou University, Zhengzhou, China}}
    \IEEEauthorblockA{\IEEEauthorrefmark{4}\textit{Communication and Information Engineering, Nanjing University of Posts and Telecommunications, Nanjing, China}}
    \IEEEauthorblockA{\IEEEauthorrefmark{5}\textit{School of Computer Science, Zhengzhou University of Aeronautics, Zhengzhou, China}}
    \IEEEauthorblockA{\{whuang, wning, ppfeng, hywang\}@ha.edu.cn , \{zmwang, iebzhou\}@zzu.edu.cn}
    \thanks{* Corresponding author.}
}

\maketitle

\begin{abstract}
Electrocardiograms (ECG), which record the electrophysiological activity of the heart, have become a crucial tool for diagnosing these diseases. In recent years, the application of deep learning techniques has significantly improved the performance of ECG signal classification. Multi-resolution feature analysis, which captures and processes information at different time scales, can extract subtle changes and overall trends in ECG signals, showing unique advantages. However, common multi-resolution analysis methods based on simple feature addition or concatenation may lead to the neglect of low-resolution features, affecting model performance. To address this issue, this paper proposes the Multi-Resolution Mutual Learning Network (MRM-Net). MRM-Net includes a dual-resolution attention architecture and a feature complementary mechanism. The dual-resolution attention architecture processes high-resolution and low-resolution features in parallel. Through the attention mechanism, the high-resolution and low-resolution branches can focus on subtle waveform changes and overall rhythm patterns, enhancing the ability to capture critical features in ECG signals. Meanwhile, the feature complementary mechanism introduces mutual feature learning after each layer of the feature extractor. This allows features at different resolutions to reinforce each other, thereby reducing information loss and improving model performance and robustness. Experiments on the PTB-XL and CPSC2018 datasets demonstrate that MRM-Net significantly outperforms existing methods in multi-label ECG classification performance. The code for our framework will be publicly available at \url{https://github.com/wxhdf/MRM}.
\end{abstract}


\section{Introduction}
Cardiovascular diseases are among the most common fatal illnesses. Studies estimate that in 2019, the number of deaths caused by cardiovascular diseases reached as high as 18.6 million \cite{b1}. Electrocardiograms (ECG), as a tool for recording the electrophysiological activity of the heart, have become a crucial basis for diagnosing cardiovascular diseases. Early detection of abnormal heart rhythms can significantly reduce the risk of sudden death caused by cardiovascular diseases, making rapid and accurate ECG diagnosis essential in clinical practice.

In recent years, deep learning models have made significant progress in ECG-assisted analysis \cite{b2}, \cite{b3}. Feature extraction is key to the classification of ECG signals. For instance, Li et al. \cite{b4} proposed a multi-label feature selection method , Yang et al. \cite{b5} introduced a multi-view and multi-scale deep neural network model , and Zhang et al. \cite{b6} a multi-scale deep residual network that combines 2-dimensional and 1-dimensional convolution blocks. Although these methods have enhanced ECG classification performance by improving feature extraction techniques, challenges remain in capturing both local waveform changes and global evolutionary trends of ECG signals.

To further explore the temporal complexity of local and global ECG features, researchers have been inspired by multi-resolution learning methods in the field of computer vision \cite{b7}, \cite{b8} and have applied these methods to ECG signal analysis. Multi-resolution feature analysis captures and processes information at different time scales, enabling the extraction of both subtle changes and overall trends in ECG signals. Among these approaches, Cai et al. \cite{b9} proposed multi-dilation convolutional blocks, and Gao et al. \cite{b10} introduced a novel multi-resolution architecture based on convolutional neural networks (CNN). Although these methods address multi-resolution features, they still face the following challenges. First, in multi-resolution feature extraction, how to effectively extract and utilize the key information from both local and global aspects of ECG signals. Second, in multi-resolution feature fusion, simple addition or concatenation of low-resolution and high-resolution features may result in some key features being overlooked. Therefore, further research is needed to explore more effective multi-resolution feature fusion methods to fully leverage the complementary advantages of features at different resolutions.

In response to the aforementioned issues, this paper proposes a novel Multi-Resolution Mutual Learning Network (MRM-Net). MRM-Net primarily comprises a dual-resolution attention architecture and a feature complementary mechanism. During training, the dual-resolution attention architecture processes high-resolution and low-resolution features in parallel, with fine-tuning applied to the different branches. The high-resolution branch focuses on capturing minute details of changes in ECG signals, while the low-resolution branch focuses on analyzing overall rhythm patterns. Both branches use an attention mechanism to weight features at different scales, thereby enhancing the model's ability to recognize and capture critical ECG features at different resolutions. To fully exploit the complementary advantages of different resolution features, MRM-Net implements knowledge interaction through a feature complementary mechanism. Specifically, after each layer of the feature extractor, we introduce a feature mutual learning module that alternately updates high-resolution and low-resolution features to achieve information sharing. This mechanism not only captures fine-grained and coarse-grained information but also achieves deep fusion at the feature level, reducing information loss. Compared to existing networks, MRM-Net demonstrates the following advantages in ECG classification:

\begin{itemize}
\item To enhance the recognition and capture of critical features, we employ a dual-resolution attention architecture that processes high-resolution and low-resolution features in parallel. The attention mechanism enables the high-resolution and low-resolution branches to focus on subtle changes and overall patterns, respectively, thereby improving the capture of critical features.
\item To effectively integrate the advantages of different resolution features, we adopt a mutual learning mechanism for the collaborative optimization of multi-resolution features. This mechanism aligns features after each layer of the feature extractor, integrating multi-resolution information. This allows features at different resolutions to reinforce each other, reducing information loss and enhancing the performance and robustness of the classification model.
\item Experimental results on the PTB-XL and CPSC2018 benchmark datasets demonstrate that MRM-Net significantly outperforms existing methods in ECG classification performance, proving its potential for practical applications.
\end{itemize}

\begin{figure*}[h]
\centerline{\includegraphics[width=1.0\textwidth]{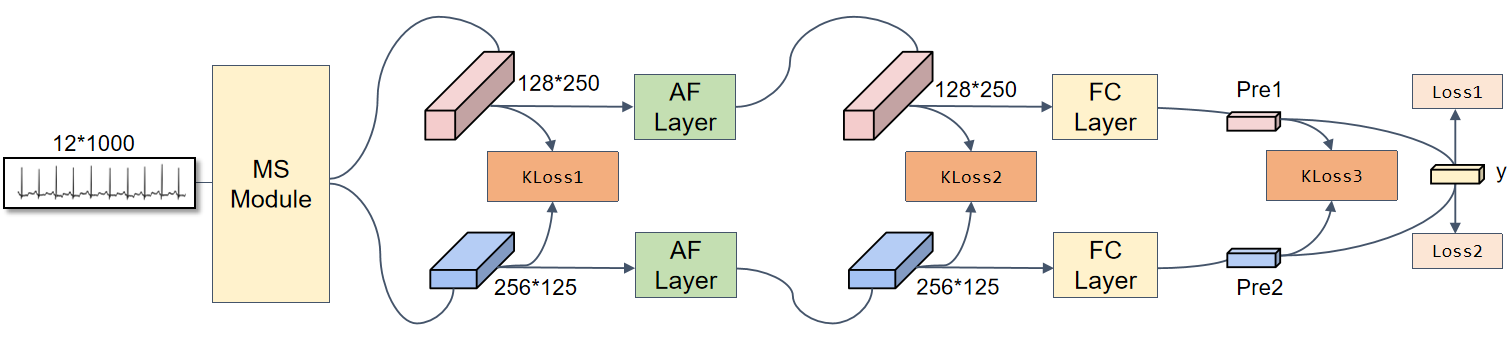}}
\caption{The overall architecture of MRM-Net comprises three main modules: the Multi-Scale Convolution Module (MS Module), the Dual-Resolution Attention Module, and the Feature Complementary Module. The Dual-Resolution Attention Module includes an Attention Fusion Layer (AF Layer) and a Fully Connected Classification Layer (FC Layer). The Feature Complementary Module contains three KLoss layers. Here, y represents the true labels, while Pre1 and Pre2 are the predictions from the low-resolution and high-resolution branches, respectively.}
\label{fig}
\end{figure*}

\section{RELATED WORKS}

\subsection{Multi-Label ECG Classification}

In recent years, deep learning-based ECG diagnostic methods have made significant progress, providing strong support for the accurate diagnosis and timely treatment of arrhythmias. Researchers have proposed various deep learning methods to improve the accuracy and efficiency of heartbeat classification. For example, Yang et al. \cite{b5} developed a multi-view, multi-scale deep neural network model for ECG classification, improving performance significantly. Pu et al. \cite{b11} proposed an innovative multi-label 12-lead ECG classification method that effectively addresses class imbalances. Xie et al. \cite{b12} introduced Hygeia, a deep learning-based method for analyzing and classifying 55 types of ECGs, enhancing performance and accuracy. Zhao et al. \cite{b12.5} proposed a novel model, ECGNN, which effectively leverages the relationships between ECG leads by combining graph convolution and graph pooling modules, significantly enhancing the accuracy and scalability of automated ECG abnormality detection.

\subsection{Mutual Learning}

Mutual learning \cite{b13} was initially used for sharing information between multiple models, allowing them to exchange information during training. This method achieves knowledge complementarity and sharing through collaborative learning among models, thereby enhancing overall system performance. The basic idea of mutual learning is to enable multiple models to guide and provide feedback to each other during training, thereby improving their respective learning outcomes. This approach has achieved significant results not only in the field of computer vision \cite{b14}\cite{b15}\cite{b16} but is also gradually being applied to one-dimensional physiological signals. For example, Ye et al. \cite{b17} proposed MLBNet, which achieved significant performance advantages in EEG emotion recognition by collaboratively training time-biased and space-biased feature learners, outperforming existing models on the DEAP dataset. Additionally, Lin et al. \cite{b18} developed a mutual learning system that dynamically updates deep learning classifier parameters, stabilizing users' EEG patterns during motor imagery and attention tasks. However, the potential of mutual learning has yet to be fully explored and utilized in the field of ECG.

\section{Method}
\subsection{Network overview}

The overall architecture of our proposed network, MRM-Net, is shown in Figure 1. It primarily consists of a multi-scale convolution module, a dual-resolution attention module, and a feature complementary module. The multi-scale convolution module is responsible for capturing the spatial features of ECG signals and adapting to their scale variations. The dual-resolution attention module includes an attention fusion layer and a linear classification layer, which process high-resolution and low-resolution features in parallel and enhance the model's focus on important information by optimizing feature representations. Additionally, the feature complementary module includes three KLoss layers. Each time a new feature is generated by a module, feature complementary performs mutual learning among the features to achieve collaborative optimization of features at different resolutions, thereby mutually enhancing the high-resolution and low-resolution branches. In the following sections, we will describe each module in detail.

\begin{table}[h]
\centering
\caption{Convolutional Layer Parameters of the Multi-Scale Convolution Module. In Ch. stands for Input Channels, and Out Ch. stands for Output Channels.}
\begin{tabular}{lccccc}
\hline
\textbf{Layer} & \textbf{Filter} & \textbf{Stride} & \textbf{Padding} & \textbf{In Ch.} & \textbf{Out Ch.} \\
\hline
Conv1 & 11 & 1 & 5 & 12 & 32 \\
Conv2 & 7 & 1 & 3 & 32 & 64 \\
Conv3 & 5 & 1 & 2 & 64 & 128 \\
Conv4 & 3 & 1 & 1 & 128 & 128 \\
\hline
\end{tabular}
\label{table1}
\end{table}

\subsection{Multi-Scale Convolution Module}\label{Multi-Scale Convolution Module}
\begin{figure}[h]
\centerline{\includegraphics[width=0.4\textwidth]{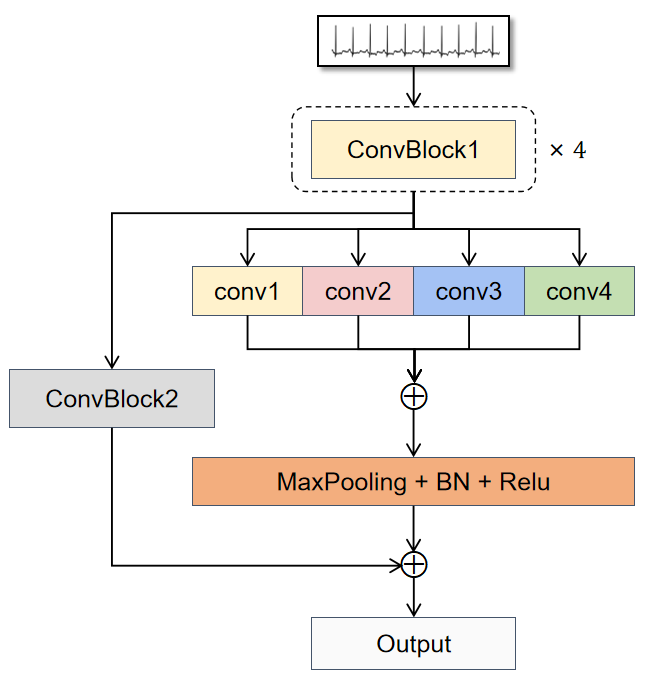}}
\caption{Schematic diagram of the Multi-Scale Convolution Module. ConvBlock consists of a convolution layer, a BatchNorm1d layer, a ReLU activation function, and a MaxPool1d layer. conv denotes the convolution layer, and $\oplus$ denotes element-wise addition.}
\label{fig2}
\end{figure}

The design of the Multi-Scale Convolution Module is inspired by the InceptionTime \cite{b19} architecture, which is suitable for capturing the spatial features of ECG signals and adapting to their scale variations. Specifically, this module includes four convolution blocks and a multi-scale fusion block, as shown in Figure \ref{fig2}. Each convolution block (ConvBlock1) consists of a convolution layer, a BatchNorm1d layer, a ReLU activation function, and a MaxPool1d layer. The parameters of the convolution layers in the convolution blocks are listed in Table \ref{table1}. The kernel size of MaxPool1d is 2, with a stride and padding of 1. In the multi-scale fusion block, four convolution layers with different kernel sizes (3, 13, 23, 33) are set up, to achieve feature fusion through weighted averaging. The fused features undergo batch normalization, ReLU activation, dropout processing, and max pooling, further integrating and compressing the information. Additionally, we designed a ConvBlock2 to process features in parallel, enabling the fusion of features from the basic convolution block and the multi-scale convolution layers, thereby enhancing the model's representation capability for the signals.
In summary, we define the ECG signal as $X \in \mathbb{R}^{m \times n}$, where $m$ is the number of leads and $n$ is the signal length. To obtain different resolution features $Z_1$ and $Z_2$, the ECG signal $X$ is processed in parallel through four convolution blocks and a multi-scale fusion block. For $Z_1$, all convolution layers in the multi-scale fusion block have an output channel count of $C$ (where $C$ is 128), with a stride of 1. For $Z_2$, all convolution layers in the multi-scale fusion block have an output channel count of $2C$ and a stride of 2. This results in two features of different resolutions, $Z_1 \in \mathbb{R}^{C \times 2N}$ and $Z_2 \in \mathbb{R}^{2C \times N}$.

\begin{figure*}[h]
\centerline{\includegraphics[width=1.0\textwidth]{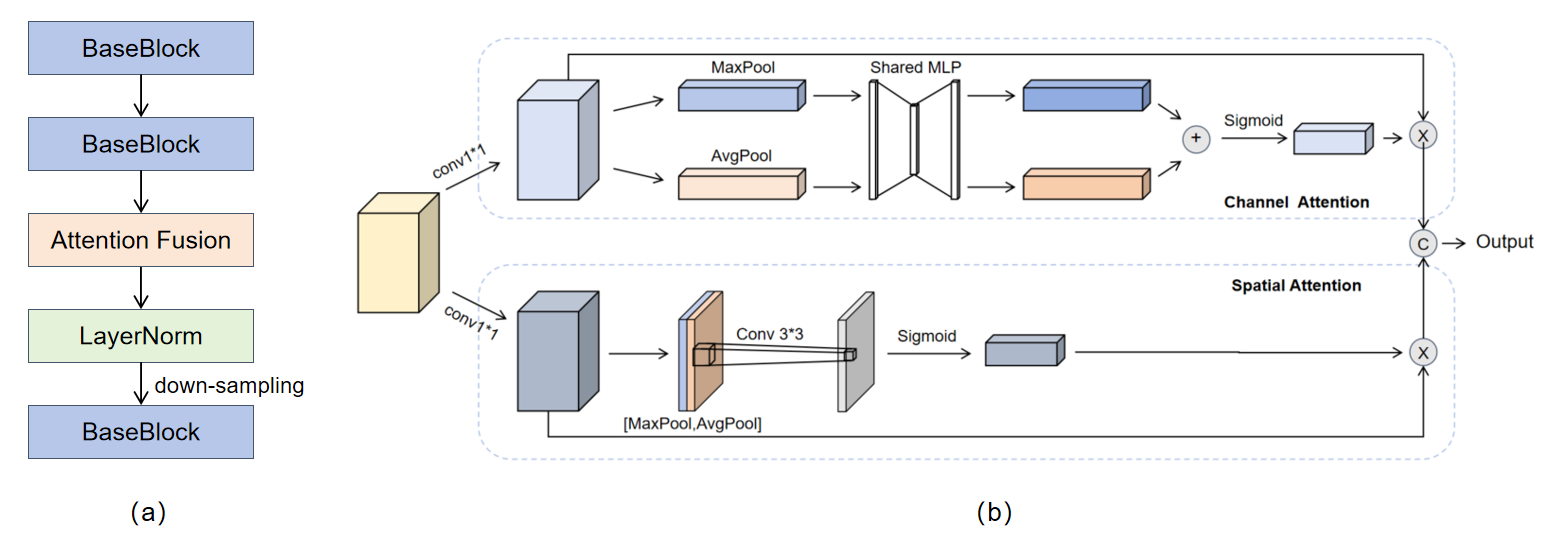}}
\caption{(a) Illustration of the Attention Fusion Layer, which includes three BaseBlocks, an Attention Fusion layer, and a LayerNorm layer. (b) Detailed illustration of the Attention Fusion. In this diagram, BaseBlock represents a residual block, conv denotes the convolution layer, AvgPool and MaxPool represent global average pooling and global max pooling, respectively, $\oplus$ denotes element-wise addition, $\otimes$ denotes element-wise multiplication, and $\textcircled{c}$ denotes the concatenation operation.}
\label{fig3}
\end{figure*}

\subsection{Dual-Resolution Attention Module}\label{Dual-Resolution Attention Module}
The Dual-Resolution Attention Module includes an Attention Fusion Layer and a Linear Classification Layer. The Attention Fusion Layer consists of three BaseBlocks, an Attention Fusion layer, and a LayerNorm layer, as shown in Figure \ref{fig3}. A BaseBlock is a residual block that contains two convolution layers, two batch normalization layers, and a residual connection, which helps mitigate the problems of vanishing and exploding gradients.In the Attention Fusion layer, features weighted by channel attention \cite{b20} and spatial attention \cite{b21} are fused, allowing the network to fully utilize attention information from different dimensions. Specifically, the channel attention mechanism focuses on the importance of different channels in the feature map, enhancing the network's attention to important channel information. The spatial attention mechanism, on the other hand, focuses on the importance of different spatial positions in the feature map, improving the network's perception of key spatial locations. Through this fusion strategy, the Attention Fusion layer not only captures richer feature representations but also enhances feature expression capabilities across multiple scales and dimensions, thereby improving the overall performance of the network and its adaptability to complex scenarios.

Taking $Z_1$ obtained from the multi-scale convolution module as an example, the formulas for the two types of attention are as follows:
\begin{align}
\text{CA}(Z_1) = \sigma\left(\text{MLP}(\text{Avg}(Z_1)) + \text{MLP}(\text{Max}(Z_1))\right)
\end{align}

\begin{align}
\text{SA}(Z_1) = \sigma\left(Conv_{3\times3}\left(\text{Cat}\left[\text{Avg}(Z_1), \text{Max}(Z_1)\right]\right)\right)
\end{align}

Where Avg denotes global average pooling, Max denotes global max pooling, Conv3×3 represents convolution operations, and $\sigma$ is the Sigmoid activation function. The two data are then concatenated through cat and downsampled through BaseBlock, with the specific operations as follows:

\begin{align}
\text{SA}(Z_1) = \sigma\left(Conv_{3\times3}\left(\text{Cat}\left[\text{Avg}(Z_1), \text{Max}(Z_1)\right]\right)\right)
\end{align}

Where LN stands for Layer Normalization, used to normalize the concatenated results, and Concat[] denotes the concatenation of the outputs from the channel attention and spatial attention mechanisms, respectively multiplied by $X$. $\otimes$ denotes element-wise multiplication. $Z_1$ obtains new feature values $Z_3$ through the Attention Fusion layer. Similarly, $Z_2$ obtains new feature values $Z_4$ through the Attention Fusion layer.

The Linear Classification Layer first passes the feature values $Z_3$ and $Z_4$ through avgpool to reduce the size of the feature maps while retaining key information. Then, Layer normalization is applied to regularize and accelerate the training process, improving the model's generalization ability. Finally, a probability score is generated for each class as the model's predicted outputs, out1 and out2. Taking $Z_3$ as an example, the process of generating out1 is as follows:

\begin{align*}
\text{out}_1 = \text{Linear}\left(\text{LN}\left(\text{AvgPool}({Z_3})\right)\right)
\end{align*}

Where LN stands for Layer Normalization, and the loss function used for training the Linear layer is BCEWithLogitsLoss. The losses for out1 and out2 are calculated with respect to the true labels as follows:

\begin{align}
\mathcal{L}_d = \text{BCEWithLogitsLoss}(\text{out}_1, Y) \nonumber \\
+ \text{BCEWithLogitsLoss}(\text{out}_2, Y)
\end{align}

\begin{figure}[h]
\centerline{\includegraphics[width=0.5\textwidth]{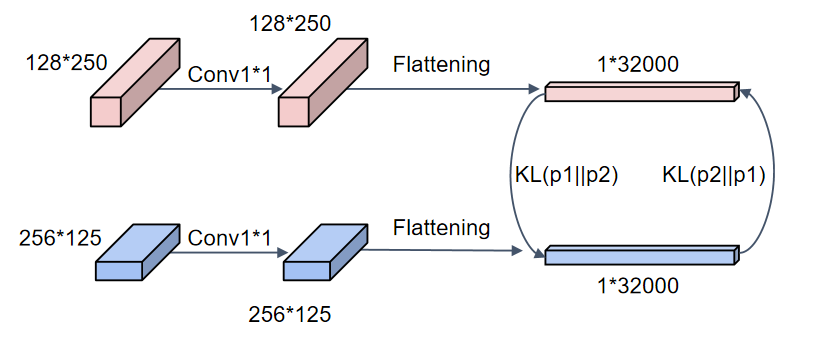}}
\caption{Diagram illustrates the feature complementary module, where Conv denotes the convolution operation, Flattening represents the flattening operation, and KL is the KL-divergence, enabling mutual learning of features at different resolutions through imitation loss.}
\label{fig4}
\end{figure}

\subsection{Feature Complementary Module}

MRM-Net employs a multi-resolution mutual learning strategy, which is different from previous networks. The two branches of MRM-Net process high-resolution and low-resolution features separately. Low-resolution features capture global information (such as overall shape and structure), while high-resolution features capture local details (such as edges and fluctuations). By using multi-resolution mutual learning, these features can be comprehensively utilized to improve the model's representation capability, as shown in Figure \ref{fig4}. During training, features of different resolutions undergo mutual learning through mimicry loss (based on KL-divergence). This mechanism encourages the two branches to learn from each other's features, facilitating the comprehensive use of global information and local details. The specific formula is as follows:

\begin{equation}
\begin{aligned}
\mathcal{L}_{\text{m}}(\text{F}_L,\text{F}_H) &= \mathcal{D}_{\text{KL}}(\text{fla}(\text{F}_L) \| \text{fla}(\text{F}_H)) \\
&\quad + \mathcal{D}_{\text{KL}}(\text{fla}(\text{F}_H) \| \text{fla}(\text{F}_L))
\end{aligned}
\end{equation}

\begin{align}
\mathcal{D}_{\text{KL}}(\text{F}_H \| \text{F}_L) &= \text{F}_H \log \frac{\text{F}_H}{\text{F}_L}, \quad 
\mathcal{D}_{\text{KL}}(\text{F}_L \| \text{F}_H) = \text{F}_L \log \frac{\text{F}_L}{\text{F}_H}
\end{align}

where $F_L$ represents low-resolution features, $F_H$ represents high-resolution features, and $\text{fla}()$ denotes the flattening operation. We perform mutual learning for the features generated by each module $(Z_1, Z_2)$, $(Z_3, Z_4)$, and $(\text{out1}, \text{out2})$. In summary, the final loss formula is as follows:

\begin{align}
\mathcal{L}_{\text{t}} &= \mathcal{L}_d + \alpha \mathcal{L}_{\text{m}}(\text{Z}_1,\text{Z}_2) + \beta \mathcal{L}_{\text{m}}(\text{Z}_3,\text{Z}_4)
+ \gamma \mathcal{L}_{\text{m}}(\text{out}_1,\text{out}_2)
\end{align}

where $\alpha$, $\beta$, and $\gamma$ are weight factors. The final loss function integrates the classification loss and the multi-resolution mutual learning loss, thus enhancing the global and local representation capabilities of the features while ensuring classification performance.

\section{EXPERIMENTS}

\subsection{Experimental Setup}
This study evaluates the performance of the proposed model using the PTB-XL Dataset \cite{b22} and the CPSC 2018 Dataset \cite{b23}. The PTB-XL dataset contains 21,837 10-second electrocardiogram (ECG) records from 18,885 patients, with a sampling frequency of 100 Hz. According to the recommended grouping method, the dataset is divided into a training set groups 1 to 8, a validation set group 9, and a test set group 10. The original sampling frequency of the CPSC 2018 dataset is 500 Hz, which is downsampled to 100 Hz to meet the experimental requirements. Signals longer than 10 seconds are truncated, while those shorter than 10 seconds are upsampled to a uniform length of 10 seconds. After preprocessing, the dataset is also divided into 10 groups, with group 9 as the validation set and group 10 as the test set.

For performance evaluation, the area under the curve (AUC), F1 score, and accuracy are selected as the main evaluation metrics. The proposed method is compared with a series of current benchmark networks and state-of-the-art methods, including fcn\_wang \cite{b24}, resnet1d\_wang \cite{b24}, MobileNetV3 \cite{b25}, InceptionTime \cite{b19}, Xresnet1d101 \cite{b26}, ECG\_BNN \cite{b11}, and MVMSNet \cite{b5}.

All networks are implemented based on the PyTorch framework. The experimental parameters are set as follows: $\alpha=0.01$, $\beta=0.01$, $\gamma=0.1$, the number of low-resolution channels is 128, and the number of high-resolution channels is 256. The batch size is 64, and the learning rate is fixed at 0.001, using the Adam optimizer. The training duration is 100 epochs. The loss function is the binary cross-entropy (BCE) loss function. All experiments are conducted on a desktop computer equipped with an Intel Core i7-12700K CPU, 32 GB RAM, and an NVIDIA RTX 3080 GPU with 12 GB VRAM. Multiple experiments are performed on each dataset, and the average metrics and corresponding standard deviations are recorded for performance comparison.

\begin{table*}[ht]
\centering
\caption{Comparison of MACRO-AUC, Accuracy, and F1 scores (mean ± standard deviation) of various deep neural networks and our network on the PTB-XL dataset for the all, diag, and sub-diag tasks. Bold indicates the best results, and underline indicates the second-best results.}
\begin{tabularx}{\textwidth}{l*{9}{c}}
\toprule
\textbf{Method} & \multicolumn{3}{c}{\textbf{all}} & \multicolumn{3}{c}{\textbf{diag}} & \multicolumn{3}{c}{\textbf{sub-diag}} \\
\cmidrule(lr){2-4} \cmidrule(lr){5-7} \cmidrule(lr){8-10}
 & \textbf{AUC} & \textbf{Accuracy} & \textbf{F1} & \textbf{AUC} & \textbf{Accuracy} & \textbf{F1} & \textbf{AUC} & \textbf{Accuracy} & \textbf{F1} \\
\midrule
fcn\_wang\cite{b24} & 88.92±1.85 & 97.71±0.12 & 70.39±0.71 & 91.11±0.78 & 97.81±0.05 & 64.85±1.89 & 91.96±0.70 & 96.15±0.09 & 67.48±0.74 \\
resnet1d\_wang\cite{b24} & 91.48±0.99 & 97.88±0.07 & 70.68±1.04 & \underline{93.01±1.08} & 97.92±0.11 & 64.51±1.32 & \underline{92.83±0.36} & 96.40±0.06 & 66.66±1.24 \\
Xresnet1d101\cite{b26} & 90.78±1.87 & 97.77±0.10 & 71.27±0.87 & 91.98±1.12 & 97.91±0.05 & 64.56±2.02 & 91.35±0.68 & 96.32±0.08 & 68.06±1.41 \\
ECG\_BNN\cite{b11} & 84.62±1.24 & 97.41±0.02 & 62.29±1.56 & 85.41±2.02 & 97.61±0.08 & 51.85±1.62 & 86.32±0.79 & 95.63±0.12 & 57.93±1.33 \\
MobileNetV3\cite{b25} & 89.04±0.77 & 97.79±0.05 & 69.91±0.86 & 88.46±3.72 & 97.49±0.07 & 61.13±1.36 & 89.53±1.97 & 96.03±0.09 & 64.35±2.21 \\
InceptionTime\cite{b19} & 91.07±1.05 & 97.83±0.11 & 71.69±0.92 & 92.81±0.94 & \underline{97.94±0.10} & 65.09±1.82 & 92.32±0.59 & 96.30±0.11 & 67.29±1.53 \\
MVMSNet\cite{b5} & \underline{92.46±0.22} & \underline{97.91±0.05} & \underline{72.43±0.67} & 92.92±0.94 & 97.92±0.06 & \underline{66.09±2.02} & 92.77±0.89 & \textbf{96.51±0.09} & \underline{69.42±1.09} \\
MRM-Net(ours) & \textbf{93.59±0.19} & \textbf{97.96±0.04} & \textbf{73.74±0.59} & \textbf{94.16±0.18} & \textbf{97.95±0.08} & \textbf{69.02±1.12} & \textbf{93.63±0.15} & \underline{96.50±0.06} & \textbf{70.59±0.96} \\
\bottomrule
\end{tabularx}
\label{table2}
\end{table*}

\begin{table*}[ht]
\centering
\caption{Comparison of MACRO-AUC, Accuracy, and F1 scores (mean ± standard deviation) of various deep neural networks and our network on the PTB-XL dataset for the Sup-diag, form, and rhythm tasks. Bold indicates the best results, and underline indicates the second-best results.}
\begin{tabularx}{\textwidth}{l*{9}{c}}
\toprule
\textbf{Method} & \multicolumn{3}{c}{\textbf{sup-diag}} & \multicolumn{3}{c}{\textbf{form}} & \multicolumn{3}{c}{\textbf{rhythm}} \\
\cmidrule(lr){2-4} \cmidrule(lr){5-7} \cmidrule(lr){8-10}
 & \textbf{AUC} & \textbf{Accuracy} & \textbf{F1} & \textbf{AUC} & \textbf{Accuracy} & \textbf{F1} & \textbf{AUC} & \textbf{Accuracy} & \textbf{F1} \\
\midrule
fcn\_wang\cite{b24} & 91.92±0.33 & 88.21±0.32 & 76.07±1.20 & 82.09±3.45 & 94.19±0.20 & 46.85±4.77 & 90.53±2.61 & 97.81±0.11 & 87.14±2.55 \\
resnet1d\_wang\cite{b24} & 91.82±0.52 & 88.29±0.71 & 75.76±2.11 & 86.88±2.01 & \underline{94.33±0.15} & 49.69±3.95 & 92.88±3.05 & 98.37±0.07 & 89.13±0.67 \\
Xresnet1d101\cite{b26} & 92.02±0.66 & 88.50±0.50 & 75.96±1.60 & 83.08±2.31 & 94.07±0.34 & 48.77±2.75 & 95.82±0.31 & 98.37±0.07 & 89.06±0.45 \\
ECG\_BNN\cite{b11} & 87.53±0.41 & 84.95±0.22 & 67.69±0.89 & 78.21±2.31 & 93.99±0.07 & 40.75±3.51 & 95.02±0.67 & 98.40±0.08 & 89.65±0.49 \\
MobileNetV3\cite{b25} & 90.60±1.07 & 87.83±0.72 & 74.58±2.32 & 80.21±4.04 & 92.95±0.52 & 45.01±4.87 & \underline{96.01±0.49} & \underline{98.50±0.09} & \underline{90.74±0.39} \\
InceptionTime\cite{b19} & 92.52±0.31 & 88.30±0.47 & 76.37±1.13 & 84.81±2.04 & 92.95±0.37 & 45.46±3.79 & 96.00±1.02 & 98.42±0.11 & 90.21±0.61 \\
MVMSNet\cite{b5} & \textbf{92.96±0.23} & \underline{89.02±0.56} & \underline{77.35±1.27} & \underline{87.42±1.82} & 94.05±0.21 & \underline{50.24±3.04} & 94.61±0.85 & 98.40±0.12 & 89.69±0.54 \\
MRM-Net(ours) & \underline{92.77±0.25} & \textbf{89.03±0.41} & \textbf{77.66±1.04} & \textbf{89.51±1.12} & \textbf{94.50±0.11} & \textbf{52.07±2.64} & \textbf{97.61±0.23} & \textbf{98.51±0.09} & \textbf{91.04±0.34} \\
\bottomrule
\end{tabularx}
\label{table3}
\end{table*}

\subsection{Classification performance}
The PTB-XL dataset contains 71 labels, subdivided into 44 diagnostic labels, 12 rhythm labels, and 19 morphology labels. These labels are further divided into 5 superclasses and 24 subclasses based on the hierarchical structure of diagnoses. Tables \ref{table2} and \ref{table3} present the results of different models on six classification tasks in the PTB-XL dataset. As shown in the tables, MRM-Net demonstrates significant performance advantages across all six classification tasks. Notably, for rhythm classification and morphology classification tasks, MRM-Net improves the AUC by 2.09\% and 1.60\% respectively, and the F1 score by 1.83\% and 0.30\% respectively, compared to the next best model.

\begin{figure}[h]
\centerline{\includegraphics[width=0.50\textwidth]{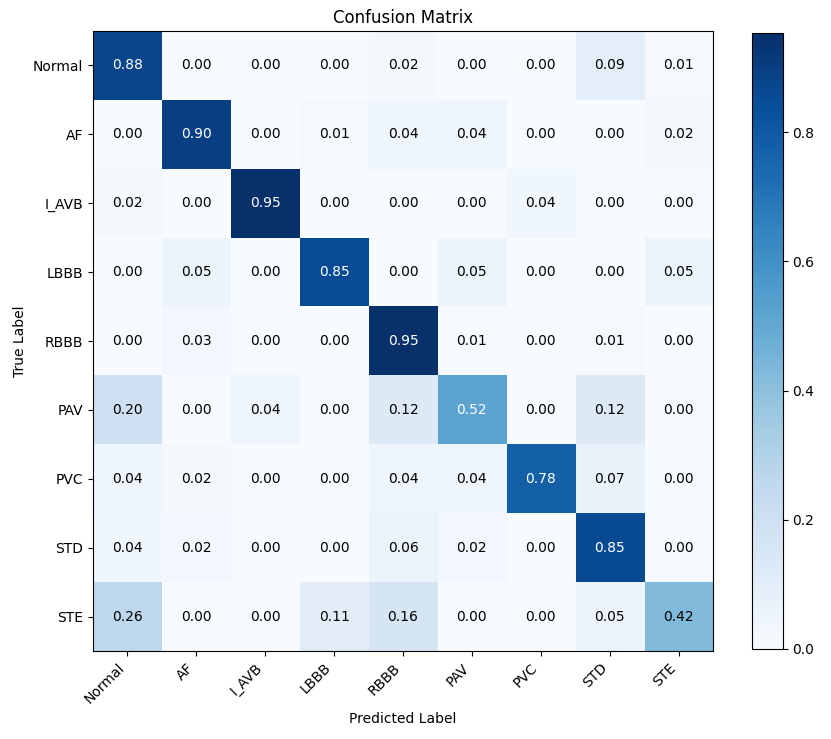}}
\caption{The confusion matrix of our method, with the horizontal axis representing the true classes and the vertical axis representing the classes predicted by our method. The values are normalized by the total number of true labels.}
\label{fig5}
\end{figure}

The CPSC 2018 dataset contains 9 categories. Table \ref{table4} presents the classification results of different models on the CPSC 2018 dataset. As shown in the table, MRM-Net outperforms other models on all three metrics. The confusion matrix of our model's classification results is shown in Figure \ref{fig5}. MRM-Net achieves excellent classification performance for most categories. However, the classification performance for STE and PAV is relatively poor, a problem also observed in other studies, likely due to the severe imbalance in sample data.

\subsection{Ablation Study}
To quantitatively analyze the classification performance of the mutual learning method in multi-resolution feature fusion, we conducted an ablation study and analyzed the results on the all task of the PTB-XL dataset and the CPSC 2018 dataset. We compared the proposed network with the feature addition and concatenation methods described in \cite{b10}. Specifically, we upsampled the low-resolution features and then either added or concatenated them with the high-resolution features. The comparison results are shown in Table \ref{table5}, where F-Addition represents feature addition, F-Concat represents concatenation followed by downsampling, and MRM-Net(ours) represents multi-resolution feature fusion through the mutual learning method.

\begin{table}[ht]
\centering
\caption{Comparison of MACRO-AUC, Accuracy, and F1 scores (mean ± standard deviation) of various deep neural networks and our network on the CPSC 2018 dataset classification tasks. Bold indicates the best results, and underline indicates the second-best results.}
\begin{tabularx}{0.5\textwidth}{l*{3}{X}}
\toprule
\textbf{Method} & \multicolumn{3}{c}{\textbf{CPSC}} \\
\cmidrule(lr){2-4}
 & \textbf{AUC} & \textbf{Accuracy} & \textbf{F1} \\
\midrule
fcn\_wang\cite{b24} & 91.68±1.69 & 94.09±0.52 & 63.84±1.60 \\
resnet1d\_wang\cite{b24} & 94.14±1.10 & 94.86±0.44 & 72.38±1.16 \\
Xresnet1d101\cite{b26} & 93.89±1.51 & 94.74±0.27 & 73.33±0.26 \\
ECG\_BNN\cite{b11} & 90.12±1.06 & 93.60±0.47 & 60.22±4.46 \\
MobileNetV3\cite{b25} & 93.29±1.19 & 94.76±0.19 & 74.53±1.39 \\
InceptionTime\cite{b19} & 94.14±0.74 & 94.90±0.29 & 73.93±2.34 \\
MVMSNet\cite{b5} & \underline{94.34±1.24} & \underline{95.13±0.14} & \underline{75.19±1.07} \\
MRM-Net(ours) & \textbf{96.14±0.17} & \textbf{95.92±0.09} & \textbf{76.24±0.97} \\
\bottomrule
\end{tabularx}
\label{table4}
\end{table}

\begin{table*}[ht]
\centering
\caption{Comparison of different feature fusion methods with our network on various evaluation metrics for the PTB-XL all task and the CPSC dataset. Bold indicates the best result, and underline indicates the second-best result. F-Addition represents feature addition, and F-Concat represents feature concatenation.}
\begin{tabularx}{\textwidth}{l*{6}{X}}
\toprule
\textbf{Method} & \multicolumn{3}{c}{\textbf{PTBXL-all}} & \multicolumn{3}{c}{\textbf{CPSC}} \\
\cmidrule(lr){2-4} \cmidrule(lr){5-7}
 & \textbf{AUC} & \textbf{Accuracy} & \textbf{F1} & \textbf{AUC} & \textbf{Accuracy} & \textbf{F1} \\
\midrule
F-Addition & \underline{93.23±0.27} & \underline{97.73±0.06} & \underline{71.50±1.35} & 95.23±0.34 & \underline{95.40±0.18} & \textbf{77.14±1.54} \\
F-Concat & 92.40±0.47 & 95.31±0.35 & 63.59±2.31 & 95.33±0.17 & 91.18±0.79 & 73.11±3.45 \\
MRM-Net(ours) & \textbf{93.59±0.19} & \textbf{97.96±0.04} & \textbf{73.74±0.59} & \textbf{96.14±0.17} & \textbf{95.92±0.09} & \underline{76.24±0.97} \\
\bottomrule
\end{tabularx}
\label{table5}
\end{table*}

\begin{table*}[ht]
\centering
\caption{Comparison of different resolution branches with our network on various evaluation metrics for the PTB-XL all task and the CPSC 2018 dataset. Bold indicates the best result, and underline indicates the second-best result. Low-RS denotes the low-resolution branch, while High-RS denotes the high-resolution branch.}
\begin{tabularx}{\textwidth}{l*{6}{X}}
\toprule
\textbf{Method} & \multicolumn{3}{c}{\textbf{PTBXL-all}} & \multicolumn{3}{c}{\textbf{CPSC}} \\
\cmidrule(lr){2-4} \cmidrule(lr){5-7}
 & \textbf{AUC} & \textbf{Accuracy} & \textbf{F1} & \textbf{AUC} & \textbf{Accuracy} & \textbf{F1} \\
\midrule
Low-RS & 92.07±0.46 & 97.84±0.06 & 71.89±0.87 & 95.51±0.29 & \underline{95.67±0.08} & \textbf{77.09±1.11} \\
High-RS & \underline{93.22±0.47} & \underline{97.92±0.03} & \underline{72.46±0.71} & \underline{95.54±0.17} & 95.53±0.12 & 76.08±1.34 \\
MRM-Net(ours) & \textbf{93.59±0.19} & \textbf{97.96±0.04} & \textbf{73.74±0.59} & \textbf{96.14±0.17} & \textbf{95.92±0.09} & \underline{76.24±0.97} \\
\bottomrule
\end{tabularx}
\label{table6}
\end{table*}

The results indicate that MRM-Net outperforms the other two feature fusion methods across most metrics. Specifically, the feature addition method exhibits the second-best performance on most metrics, while the feature concatenation method shows the worst performance across all metrics. This suggests that simple feature concatenation followed by downsampling is less effective for feature fusion in these tasks. The data in Table \ref{table5} further validate the effectiveness of our proposed mutual learning method in multi-resolution feature fusion, significantly enhancing classification performance.

Additionally, to analyze the effectiveness of the dual-branch mutual learning structure, we conducted further ablation experiments on the PTB-XL dataset's all task and the CPSC 2018 dataset. Specifically, we compared the cases of using only low-resolution or high-resolution features, where the low-resolution dimension is 128 and the high-resolution dimension is 256. The comparison results are shown in Table  \ref{table6}, where Low-RS represents using only the low-resolution branch , High-RS represents using only the high-resolution branch, and MRM-Net(ours) represents multi-resolution feature fusion through the mutual learning method.

The results show that the dual-branch mutual learning structure achieves the best results in most metrics. In the PTB-XL dataset, the high-resolution features perform better than the low-resolution features; however, in the CPSC 2018 dataset, the high-resolution features perform worse than the low-resolution features. This may be because high-resolution features contain more detailed information and perform better in more complex classification tasks, but too much irrelevant information can affect classification performance in simpler classification tasks. Overall, fusing high- and low-resolution features is currently the best choice.

\subsection{Hyperparametric Analysis}
\begin{figure}[h]
\centerline{\includegraphics[width=0.48\textwidth]{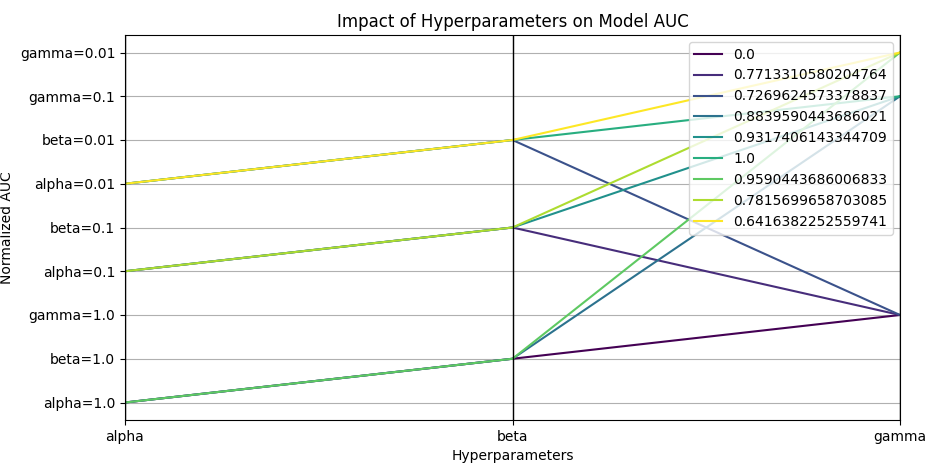}}
\caption{The impact of hyperparameters on the AUC value of the model, where alpha, beta, and gamma correspond to the hyperparameters $\alpha$, $\beta$, and $\gamma$, respectively.}
\label{fig6}
\end{figure}

This section discusses the impact of hyperparameters on model performance, specifically the selection of weight factors $\alpha$, $\beta$, and $\gamma$, as well as the selection of low-resolution and high-resolution features. To evaluate the impact of weight factors $\alpha$, $\beta$, and $\gamma$, we used parallel coordinate plots to show the effect of different configurations on model performance, as shown in Figure \ref{fig6}. The experiments used the AUC value of the CPSC 2018 dataset for evaluation.

\begin{table}[ht]
\centering
\caption{Comparison of different resolution settings on various evaluation metrics and training duration for the PTB-XL all task. Bold indicates the best result, and underline indicates the second-best result. LR-D and HR-D represent the dimensions of the low-resolution branch and high-resolution branch, respectively.
}
\begin{tabularx}{0.5\textwidth}{l*{4}{X}}
\toprule
\textbf{Method} & \multicolumn{4}{c}{\textbf{PTBXL-all}} \\
\cmidrule(lr){2-5}
 & \textbf{AUC} & \textbf{Accuracy} & \textbf{F1} & \textbf{Time} \\
\midrule
LR-D = 64, \  HR-D = 128 & 92.50±0.31 & 97.92±0.04 & 73.70±0.87 & \textbf{0m50s} \\
LR-D = 128, HR-D = 256 & \underline{93.59±0.19} & \textbf{97.96±0.04} & \underline{73.74±0.59} & 1m30s \\
LR-D = 256, HR-D = 512 & \textbf{93.87±0.15} & \underline{97.94±0.03} & \textbf{74.01±0.23} & 3m50s \\
\bottomrule
\end{tabularx}
\label{table7}
\end{table}

The experimental results indicate that smaller values of $\alpha$ and $\beta$ (such as 0.01 or 0.1) significantly improve the model's AUC compared to higher values (such as 1). Considering various configurations, the best performance is achieved with $\alpha=0.01$, $\beta=0.01$, and $\gamma=0.1$. As shown in Figure \ref{fig6}, each line represents a set of hyperparameter configurations and their impact on the model's AUC value. Different values are represented by different vertical axes, and AUC values are normalized and displayed by color intensity, with darker colors indicating higher AUC values.

For the selection of low-resolution and high-resolution features, we conducted experiments on the all task of the PTB-XL dataset, comparing three different resolution configurations: 64 and 128, 128 and 256, 256 and 512. The experimental results are shown in Table \ref{table7}. It can be seen that as the resolution increases, the AUC gradually improves, but the training time also increases correspondingly.

Specifically, when the low resolution is 64 and the high resolution is 128, the AUC reaches 92.50\% with a training time of 50 seconds. When the low resolution is 128 and the high resolution is 256, the AUC improves to 93.59\% with a training time of 1 minute and 30 seconds. When the low resolution is 256 and the high resolution is 512, the AUC further improves to 93.87\%, but the training time increases to 3 minutes and 50 seconds.

Considering model performance and training efficiency, we ultimately chose a low resolution of 128 and a high resolution of 256 as the optimal parameter configuration. This configuration achieves a reasonable balance between performance and training time, ensuring high classification accuracy while maintaining a fast training speed.

\section{CONCLUSION}
In this paper, we propose a novel Multi-Resolution Mutual Learning Network (MRM-Net). By introducing a dual-resolution attention architecture and a feature complementary mechanism, MRM-Net is able to capture critical information in ECG signals across different time scales. The dual-resolution attention architecture processes high-resolution and low-resolution features in parallel, enabling the high-resolution branch and the low-resolution branch to focus on fine changes and global patterns, respectively, through the attention mechanism, thereby enhancing the ability to capture key features. The feature complementary mechanism introduces a feature mutual learning module after each layer of the feature extractor, allowing features at each resolution to reinforce each other, reduce information loss, and thus improve the model's performance and robustness. Experimental results demonstrate that MRM-Net significantly outperforms existing methods in the multi-label classification of ECG signals on the PTB-XL and CPSC2018 benchmark datasets, proving its great potential in practical applications.

\section{ACKNOWLEDGMENT}
This work is supported by the National Natural Science Foundation of China under Grant Number 62302463, and the Natural Science Research Start-up Foundation of Recruiting Talents of Nanjing University of Posts and Telecommunications under Grant Number NY223138.

\vspace{12pt}
\color{red}

\end{document}